\documentclass[aps,nofootinbib,prd,eqsecnum,showpacs,showkeys,preprintnumbers]{revtex4-1}
\usepackage[caption=false]{subfig}
\usepackage{graphicx}
\usepackage{amsmath}
\usepackage{amsfonts}
\usepackage{amssymb}
\usepackage{color}
\usepackage{bm}
\usepackage{mathrsfs}
\usepackage{epstopdf}
\usepackage{url}
\usepackage{footnote}
\usepackage{textcomp}
\usepackage{ulem}
\usepackage{esint}
\usepackage[unicode=true, pdfusetitle,
 bookmarks=true,bookmarksnumbered=false,bookmarksopen=false,
 breaklinks=false,pdfborder={0 0 1},backref=false,colorlinks=false]{hyperref}
\usepackage{multirow}
\usepackage{pifont}

\begin{document}
\title{Constraining dark energy equations of state  in $F(R,T)$ gravity}

\author{Ahmed Errahmani$^{1}$}
\email{ahmederrahmani1@yahoo.fr}
\author{Amine Bouali$^{1,2}$}
\email{a1.bouali@ump.ac.ma}
\author{Safae Dahmani$^{1}$}
\email{dahmani.safae.1026@gmail.com}
\author{Imad El Bojaddaini$^{1}$}
\email{i.elbojaddaini@ump.ac.ma}
\author{Taoufik Ouali$^{1}$}
\email{t.ouali@ump.ac.ma}
\affiliation{$^{1}$Laboratory of Physics of Matter and Radiation, Mohammed I University, BP 717, Oujda, Morocco\\
$^{2}$ Higher School of Education and Training,\\ Mohammed I University, BP 717, Oujda, Morocco }
\date{\today}

\begin{abstract}
In this paper, we examine the acceleration of the Universe's expansion in  $F(R,T)$  gravity, where $R$ denotes the Ricci scalar and $T$ the trace of energy-momentum tensor. Indeed, the unknown nature of the source controlling this acceleration in general relativity leads scientists to investigate its properties by means of some alternative theories to general relativity.
 Our study is restricted to the particular case where 
$F(R,T)=R+2\kappa^2\lambda T$, with  $\lambda$ being a constant. We use a Bayesian analysis of  current observational datasets, including the type Ia supernovae constitution compilation and H(z) measurements, to constrain free parameters of the model. To parametrize  dark energy, we consider  two well known equations of state. We find the best fit values for each model by running a  Markov chain Monte Carlo technic. The best fit parameters are used to compare both models to $\Lambda CDM$ by means of the Akaike information criterion and the Bayesian information criterion. We show that the Universe underwent recently a transition from a deceleration to an acceleration for both models. Furthermore, the data shows a phantom nature of the equation of state for both models.

\end{abstract}

\maketitle
\date{\today}

\color{black}


\section{Introduction}\label{sec1}

 Recent cosmological observations, such as the observational discovery of the accelerated expansion of the Universe \cite{Riess,Perlmutter} have posed a major challenge to gravitational theory. To address this observational question, scientists have suggested the presence of an enigmatic type of energy density known as Dark Energy (DE) in the context of general relativity.  The current observations strongly support the  $\Lambda$CDM cosmological model  which posits that a  cosmological constant $\Lambda$ the primary source of DE and  the dark matter (DM) component is responsible for the formation of galaxies and the distribution of large-scale structures (LSS).  Despite its preference in light of current observations, the $\Lambda$CDM model grapples with two cosmological constant problems \cite{Peebles,Padmanabhan,Copeland,Frieman,Caldwell,Silvestri,Capozziello,Wang}.
 Moreover, the Hubble tension, a perplexing cosmological discrepancy, has emerged as a focal point of contemporary astrophysical inquiry. This enigmatic phenomenon stems from a persistent inconsistency in the measurements of the rate of expansion of the Universe, known as the Hubble constant. Disparate methodologies yield discordant values, leading to a fundamental discord between high redshift measurements based on $\Lambda$CDM and low redshift observations independent model. \cite{Eleonora}. The discrepancy, though subtle, carries profound implications for our understanding of the Universe's evolution, structure and ultimate fate. Addressing the Hubble tension has become an urgent imperative, compelling astronomers and cosmologists to reevaluate existing paradigms and explore new theoretical frameworks. To address these theoretical challenges, diverse alternative theories have been proposed. The first suggest the existence of non-traditional forms of energy density within the framework of General Relativity including quintessence \cite{Carroll1998,Ladghami}, phantom \cite{Caldwell,Bouali2021,Bouali2019,Mhamdi2023,dahmani2023constraining,dahmani2023smoothing,bouhmadi2015little}, k-essence \cite{Malquarti,Chiba}, holographic dark energy \cite{Horava,Li2004,Wang2017,bouhmadi2018more,belkacemi2012holographic,bargach2021dynamical,bouhmadi2011cosmology,Belkacemi2020}, and various others. The second approach involve modifying the gravitational part of Einstein’s equations in various ways. 
 Modifications of gravity are driven by the imperative to elucidate unresolved issues within the standard model of cosmology, including the accelerated expansion of the Universe and the presence of dark matter and dark energy. While the standard model of cosmology assumes the enigmatic dark energy with negative pressure fueling the accelerated expansion, its nature and origin remain elusive.
 
The extra terms contained in equations of modified gravity provide an alternative explanation for the observed cosmic acceleration without the need for dark energy.\\
The $F(R)$ theory \cite{Nojiri2003,Sotiriou,Capozziello2011} is one of the most widely explored models in which the standard Einstein-Hilbert action is replaced by an arbitrary function of the Ricci scalar and have been subject of extensive studies in recent times \cite{Wu}.\\ The first extension of $F(R)$ modified theories of gravity called $F(R,Lm)$ gravity assumes an explicit coupling of an arbitrary function of the Ricci scalar $R$ with the matter Lagrangian density $L_{m}$ \cite{Barrientos,Harko2010}.
 For an overview of the $F(R, Lm)$ gravity we refer the reader to  \cite{Harko2014}.\\ A second  extension of the Hilbert-Einstein action leads to the $F(R,T)$ class of gravitational theories \cite{Harko2011}. This theoretical framework suggests a modification to Einstein's theory of general relativity by introducing a new function of both the Ricci scalar and the trace of the energy-momentum tensor. 
 There are multiple  motivations behind this extension, such as the geometry-matter coupling contained in the trace
of the energy-momentum tensor of matter and the non conservation of the energy-momentum tensor due to quantum effects by means of the trace $T$. Furthermore, this non conservation may be interpreted as an imperfect fluid in the theory caused by anisotropy, as it appears, at high energy densities, naturally in the matter density
\cite{Ruderman, Canuto}.
 \\
The $F(R,T)$ gravity has been applied to various cosmological scenarios, including the formation and the evolution of large-scale structures in the Universe, cosmic microwave background radiation, and the growth of supermassive black holes \cite{Debabrata}. The goal is to create models that can behave similarly to a cosmological constant and attempt to explain the late-time accelerated expansion of the Universe.  Some recent studies have shown that $F(R,T)$ gravity can explain the observed cosmic acceleration without the need for dark energy and can also account for the growth of large-scale structures in the Universe, such as galaxy clusters and filaments \cite{Hamid, Hamid2, Chakraborty, Charif, NasrAhmed, Kumar}. One of the significant applications of $F(R,T)$ gravity is its ability to provide a unified explanation for the early and late-time acceleration of the Universe \cite{Bhattacharjee2022}. In particular, the theory can explain the cosmic acceleration during the inflationary epoch and at the current epoch, challenges that are difficult to reconcile within the standard model of cosmology.
Recent studies have also explored observational constraints on $F(R,T)$ modified gravity, using data from cosmic microwave background radiation, large-scale structure of the Universe, and  gravitational lensing. These studies have shown that $F(R,T)$ gravity can be consistent with observations, but the precise constraints depend on the specific form of the function $F(R,T)$ 
\cite{Kumar2017,Moraes2017PRD,Odintsov2018,Sharif2019,Chakraborty2023}.
\\ In \cite{Hamid2}, the authors discuss $F(R,T)$ models with an effective energy density and an effective equation of state parameter.
Constraining the equation of state EoS of dark energy is a different approach to studying these models. In fact, the paper's goal may be seen in this perspective. In this work, we propose two models parameterized by two equations of state in modified $F(R,T)$ gravity theory to discuss the current acceleration of the expansion.
We identify the most suitable values for each equation of state parameters by constraining the model with observational data.
Model parameters will therefore be constrained using  Pantheon Plus dataset sample, which includes 1701 points covering the redshift range $0.001 < z < 2.26$ \cite{Pan}, and the observational datasets, which are the H(z) data (denoted as OHD) with 57 data points of the Hubble parameter\cite{Sharov}. To this aim, we perform a Markov Chain Monte Carlo analysis and use the Akaike's information criterion (AIC) \cite{AIC} and the Bayesian informations criterion (BIC) \cite{BIC} to classify these models with respect to the $\Lambda$CDM model considered as a reference. \\

Combining parameterized EoS, characterizing DE, the physics content of the Universe and  the modified $f (R,T)$ gravity, our paper explores two forms of EoS. The first EoS is the well known  Chevallier-Polarski- Linder (CPL)  parametrisation  which is obtained as a Taylor series expansion, up to the first order of the scale factor.
The second  EoS parameter has a logarithmic form (see Sec. \ref{subsec3c}).  Using observational data to constrain EoS parameters and comparing them in the framework of $f(R,T)$ modified gravity is a compelling avenue of investigation.\\

In the subsequent sections of this paper, we embark on a detailed exploration of the proposed models. Section \ref{sec2} delves into the foundational equations of $F(R,T)$ gravity, providing the theoretical groundwork. Moving forward, Section \ref{sec3} introduces cosmological models characterized by parametrized equations of state and their corresponding Hubble parameters. The observational component is addressed in Section \ref{sec4}, where we detail the datasets utilized and the methodologies employed for model fitting, drawing comparisons with the $\Lambda$CDM model. The ensuing sections, \ref{sec5} and \ref{sec6}, present the main results of our study and a conclusion, respectively.
\section{$F(R,T)$ SETUP}\label{sec2} 
In the present section, we introduce the geometric aspect of the action  for the $F(R,T)$ gravity theory. Generalizing the Hilbert action, 
we assume that $F(R,T)$ gravity  is given 
by   \cite{Harko2011}

\begin{equation}
S=\int \sqrt{-g}d^{4}x\left[ \frac{1}{2\kappa^{2}}F(R,T)+\mathcal{L}_{m}%
\right],\label{S}
\end{equation}
where $g$ is the determinant of the metric tensor, $g_{\mu\nu}$, $\kappa^2 =8\pi G$ is the gravitational constant, $R$ is the
Ricci scalar curvature, $\mathcal{L}_{m}$ is the Lagrangian density of any
matter fields and $T$ is the trace of the stress-energy tensor, $T_{\mu\nu}$, defined as
\begin{equation}
T_{\mu \nu }=-\frac{2}{\sqrt{-g}}\frac{\delta (\sqrt{-g}\mathcal{L}_{m})}{%
\delta g^{\mu \nu }}=2\frac{\delta (\mathcal{L}_{m})}{\delta g^{\mu \nu }}%
-g_{\mu \nu }\mathcal{L}_{m}.
\end{equation}

The field equations are obtained by varying the action, (\ref{S}), with respect to the metric tensor. The resulting field equations are
\begin{equation}
(R_{\mu \nu }+g_{\mu \nu }\square -\nabla _{\mu }\nabla _{\nu })F_{R}(R,T)-%
\frac{1}{2}F(R,T)g_{\mu \nu }=-F_{T}(R,T)(\Theta _{\mu \nu }+T_{\mu \nu })+\kappa^2 T_{\mu \nu },
\label{Fieldeq}
\end{equation}
where $F_{R}(R,T)=\partial F(R,T)/\partial R$, $F_{T}(R,T)=\partial
F(R,T)/\partial T$, $\nabla _{\nu }$ is the covariant derivative, $\square$ is the D'Alembert operator \ and 
\begin{equation}
\Theta _{\mu \nu }=g^{\gamma \rho }\frac{\delta T_{\gamma \rho }}{\delta
g^{\mu \nu }}=g_{\mu \nu }\mathcal{L}_{m}-2T_{\mu \nu }-2g^{\gamma \rho }%
\frac{\partial ^{2}\mathcal{L}_{m}}{\partial g^{\mu \nu }\partial g^{\gamma
\rho }}.
\label{teta}
\end{equation}
One of the most intriguing properties of  $F(R,T)$ gravity is the non conservation of the energy-momentum tensor
\begin{equation}
(2\kappa^2-F_T)\nabla^\mu T _{\mu \nu }=(T _{\mu \nu }+\Theta _{\mu \nu }) \nabla^\mu F_T+F_T \nabla^\mu\Theta _{\mu \nu }+\nabla^\mu R_{\mu \nu } -\frac{1}{2}g_{\mu \nu }\nabla^\mu F.
\end{equation}
This non conservation  is perceived as a transfer of the energy and the momentum between the geometry and the matter.

To apply this modified gravity to cosmology, we consider a flat Friedmann-Lemaitre-Robertson-Walker (FLRW) line element given by

\begin{equation}
d s^2=-d t^2+a^2(t)\left(d r^2+r^2 d \Omega^2\right)
\label{4},
\end{equation}

where $a(t)$ is the scale factor, $d\Omega^2=d\theta^2+\sin\theta^2d\phi^2$ and t represents the cosmic time.
\\

The lagrangian density $L_m$ can be assumed as $L_m = -p$. 
Applying this to Eqs. \eqref{Fieldeq} and  \eqref{teta}, we obtain
\\

\begin{equation}
 3 H^2 F(\mathrm{R}, \mathrm{T})+\frac{1}{2}(f(\mathrm{R}, \mathrm{T})-F(\mathrm{R}, \mathrm{T}) \mathrm{R})+3 \dot{F}(\mathrm{R}, \mathrm{T}) H 
 =\left(\kappa^2+\mathcal{F}(\mathrm{R}, \mathrm{T})\right) \rho+\mathcal{F}(\mathrm{R}, \mathrm{T}) p,
\label{9'}
\end{equation}
\\

and

\begin{equation}
2 F(\mathrm{R}, \mathrm{T}) \dot{H}+\ddot{F}(\mathrm{R}, \mathrm{T})-\dot{F}(\mathrm{R}, \mathrm{T}) H=-\left(\kappa^2+\mathcal{F}(\mathrm{R}, \mathrm{T})\right)(\rho+p),
\label{10'}
\end{equation}
where $p$ and $\rho$ are the pressure and energy density of the cosmic fluid, respectively.\\

In the next step of the paper, we will illustrate our purpose by a specific and a simple form of $F(R,T)$ gravity, which can be written as:. 
\begin{equation}
F(R,T)=R+2\kappa^2\lambda T,
\label{model}
\end{equation}

where $\lambda$ is a free parameter.
\\


{Substituting \eqref{model} into Eqs \eqref{9'} and \eqref{10'}, the Friedmann equations can be expressed as\\
\begin{equation}
3H^{2}=\kappa ^{2}( \rho +3\lambda \rho-\lambda p),
\label{Friedmann1}
\end{equation}
\begin{equation}
-2\overset{\cdot}{H}-3H^{2}=\kappa ^{2} (p+3\lambda p-\lambda \rho) ,
\label{Friedmann2}
\end{equation}
where we have considered that the stress energy-momentum tensor describes a perfect fluid i.e.
\begin{equation}
T_{\mu \nu }=(p+\rho )u_{\mu }u_{\nu }-pg_{\mu \nu },
\end{equation}
with $u^{\mu }$ is the four velocity. 

 By expressing the energy density and the pressure
in term of the Hubble rate, Eqs. \eqref{Friedmann1} and \eqref{Friedmann2} become
\begin{equation}
\kappa ^{2}(1+4\lambda )\rho =3H^{2}-\frac{2\lambda }{2\lambda +1}%
\overset{\cdot }{H},
\label{Rho}
\end{equation}
and
\begin{equation}
\kappa ^{2}(1+4\lambda )p=-3H^{2}-2\frac{3\lambda +1}{2\lambda +%
1}\overset{\cdot }{H}.
\label{P}
\end{equation}
Furthermore, by introducing 
the energy density, $\rho_{DE}$, and 
the  pressure, $p_{DE}$,
the above Friedmann equations can be rewritten as
\begin{equation}
3H^{2}=\kappa ^{2}(\rho +\rho_{DE})
\label{FreidmannDE}
\end{equation}
\begin{equation}
-2\overset{\cdot }{H}-3H^{2}=\kappa^{2}(p+p_{DE})
\label{FreidmannDP}
\end{equation}
where 
\begin{equation}
\rho_{DE}=\lambda(3\rho-p),
\label{rhoDE}
\end{equation}
and
\begin{equation}
p_{DE}=\lambda(3p-\rho).
\label{PDE}
\end{equation}
For $\lambda=0$, we recover $\Lambda$CDM. \\

In the literature, the expression of the interaction term between cosmic fluids and dark energy densities is often formulated using phenomenological approaches, due to the absence of a definitive theory to specify its precise form. In our case, the nature of this interaction is deduced  from the non conservation of the stress energy tensor in $f(R,T)$ gravity. Indeed, using the form of $F(R,T)$, Eq. \eqref{Fieldeq} can be written as

\begin{equation}
G_{ \mu\nu} = \kappa^2 T^{eff}_{ \mu\nu},
\label{G}
\end{equation}
where \quad $T^{eff}_{ \mu\nu} = T_{\mu\nu} + \tilde{T}_{\mu\nu}$, and  $ \tilde{T}_{\mu\nu}  \equiv \lambda [2(T_{\mu\nu} + pg_{\mu\nu} ) + T g_{\mu\nu}]$.
\\

Nonetheless, when we apply the Bianchi identities, $\nabla_{ \mu}G_{ \mu\nu} =0$, to Eq. \eqref{G}, we derive

\begin{equation}
\dot{\rho}+3 H(\rho+p)=-\frac{\lambda}{\left(1+2 \lambda\right)}(\dot{\rho}-\dot{p}).
\label{rhodot}
\end{equation}
\\

 Furthermore, starting from Eqs. \eqref{rhoDE} and \eqref{PDE}, we can observe that the  dark energy density, $\rho_{DE}$,  is not conserved

\begin{equation}
\dot{\rho}_{DE}+3 H(\rho_{DE}+p_{DE})=\frac{\lambda}{\left(1+2 \lambda\right)}(\dot{\rho}-\dot{p}).
\label{rhoDEdot}
\end{equation}
\\

Furthermore, we can easily show that the two components of the budget of the Universe i.e. $\rho$ and $\rho_{DE}$, interact with each other with the interaction term $Q=\lambda (\dot{\rho}-\dot{p})/(1+2\lambda)$.  
By setting $\rho_{eff}=\rho+\rho_{DE}$, we can show from Eqs. (\ref{rhodot}) and (\ref{rhoDEdot}) that $\rho_{eff}$ fulfills the continuity equation and hence $\rho$ and $\rho_{DE}$ fulfill in turn the continuity equation with the above interaction term
\begin{equation}
 \dot{\rho}_{eff}+3H(\rho_{eff}+p_{eff})=0.   
\end{equation}

This interaction reflects the non conservation of energy and momentum i.e. between the geometry and the matter or in our consideration between DE and matter. This non conservation reflects the non-geodesic motion
of massive test particles \cite{Harko2011} or  related to an irreversible matter creation process \cite{Harko2014}.


With the aid of Eqs. (\ref{Rho}, (\ref{P}), (\ref{rhoDE}) and (\ref{PDE}) the equation of state (EoS) parameter of dark energy,  $\omega=p_{DE}/\rho_{DE}$,  writes
\begin{equation}
\omega =-\frac{H^{2}+A\overset{\cdot }{H}}{H^{2}+B\overset{\cdot }{H}},
\label{wdark}
\end{equation}
where $A=\frac{(8\lambda+3)}{6(2\lambda+1)}$ and $B=\frac{1}{6(2\lambda+1)}$.\\
Using
\begin{equation}
\frac{d}{dt}=-(1+z)H\frac{d}{dz},
\label{dtdz}
\end{equation}
Eq. (\ref{wdark}) can be rewritten as 
\begin{equation}
\frac{dH^{2}}{H^{2}}=\frac{2}{(1+z)}\frac{(1+\omega )}{(A+B\omega)}dz.\label{Hz}
\end{equation}
 To solve this equation, an ansatz of the EoS parameter of dark energy described by $F(R,T)$ gravity is required. The next section will be devoted to this aim by
considering two forms of EoS.
\section{Equations of state}\label{sec3}
The ability of $F(R,T)$ gravity models to circumvent the cosmological constant problem is well recognized in the literature. Indeed, the extra terms that appear in their field equations allow to explain the current cosmic acceleration of the Universe \cite{Moraes2018, Sing2016, Kumar2015, Barrientos2014, Moraes2017, Moares2016, Correa2016}. Furthermore, dark energy may also be expressed in terms of an effective EoS parameter. In this direction, 
we will discuss two models parameterized by  an EoS parameter which were mostly used in the literature to describe the late-time acceleration of the Universe.
\subsection{Model A}\label{subsec3a}
Investigating a possible dynamic of DE, Chevallier Polarski and Linder  introduced in \cite{Chevallier2001,Linder2003} an EoS parameter as follows
\begin{equation}
\omega  
=\omega _{0}+\omega _{1}\left(\frac{z}{1+z}\right),  
\end{equation}
where the free parameter $\omega_{0}$ denotes the current value of the equation of state and $\omega _{1}=d\omega/dz$ its derivative  at the present epoch.  While the sign of $\omega _{1}$ gives  an insight of the evolution of  the EoS parameter i.e. a decreasing or increasing with respect to the redshift, the sign of $1+\omega_0$ indicates whether the dynamical DE behaves currently as a phantom ($\omega_0<-1$) or a quintessence ($\omega_0>-1$) behaviour. This parametrization can reproduce $\Lambda$CDM for  $\omega_{0}=-1$ and $\omega_{1}=0$. Furthermore, even it diverges at the future i.e. at $z\rightarrow-1$, the Chevallier Polarski and Linder EoS converges for high redshifts and gives an  accurate reconstruction for several scalar fields.

By substituting this
equation into Eq. (2.21), we obtain the following modified Friedmann equation
\begin{equation}
\frac{H^{2}}{H_{0}^{2}}=\Big(1+z\Big)^{12(1+2\lambda )}%
\left( 1+(1+\frac{\omega_{1}}{8\lambda + (3+\omega_{0})})z\right) ^{-24%
\frac{(1+4\lambda )(1+2\lambda )}{8\lambda + (3+\omega_{0}+\omega_{1})}},
\end{equation}
where $H_{0}$ is the present value of the Hubble rate.

\subsection{Model B}\label{subsec3c}
The second  form of the EoS parameter in which we are interested,
has a logarithmic form and was introduced in \cite{Efstathiou1999} as
\begin{eqnarray}
\omega  
&=&\omega _{0}+\omega _{1}\ln (1+z),
\end{eqnarray}
where  $\omega_{0}$ and $\omega _{1}$ are the present free parameters of $\omega$ and  its derivative, respectively.  Substituting this equation into Eq. (\ref{Hz}). The modified Friedmann equation writes
\begin{equation}
\frac{H^{2}}{H_{0}^{2}}=\left( 1+\frac{\omega _{1}}{8\lambda +3+\omega_{0}}\ln (1+z)\right) ^{-24\frac{(1+4\lambda)(1 +2\lambda)}{\omega_{1}}}\Big(1+z\Big)^{12(1+2\lambda )}.
\end{equation}

\subsection{Cosmic acceleration }
In addition to the EoS parameter which characterizes the cosmological fluid, the deceleration parameter $q$ can tell us about the current and late dynamics of the Universe. The deceleration parameter is defined as
\begin{equation}
q=-1-\frac{\overset{.}{H}}{H^{2}}.
\label{q}
\end{equation}
By using Eqs. (\ref{wdark}) and (\ref{q}), the deceleration parameter can be  rewritten in terms of the EoS parameter as

\begin{equation}
q=-1+\frac{{1+\omega}}{A+B\omega}.
\label{decelerationqw}
\end{equation}

\section{Observational data and Methodology }\label{sec4}
The observational constraints described in this section will be used to estimate the model's parameters for each form of EoS. We perform the minimization of the chi-square function, $\chi^2$, using the Markov Chain Monte Carlo (MCMC) algorithm \cite{MCMC} in order to explore the parameter space. The best model is the one that corresponds to the small value of $\chi^2$. The chi-square function, for a gaussian distribution, writes $\chi^2=-2\ln{(\mathcal{L}_{max})}$ where $\mathcal{L}$ is the likelihood function. In our analysis we combine the $H(z)$ measurements \cite{Sharov} (abbreviated as OHD) with the recent Pantheon Plus dataset \cite{Pan} (abbreviated as Pantheon$^+$) in order to estimate the cosmological parameters.\\

\textbf{OHD}: we use 57 data points of the Hubble parameter of $H(z)$ measurements data  \cite{Sharov}. The chi-square of these data is given by
\begin{equation}
\chi^2_{OHD}(\alpha,\beta)=\sum_{i=1}^{57}\left[\frac{H_{th}(z_i\mid \alpha,\beta)-H_{obs}(z_i)}{%
\sigma_i}\right]^2,
\end{equation}
where $H_{th}(z_i\mid\alpha,\beta)$ is the theoretical values predicted by the model of the Hubble parameter at redshift $z_i$, $H_{obs}(z_i)$ is the observed values at the same redshift $z_i$ and $\sigma_i$ is the standard deviation. ($\alpha$, $\beta$) are the free parameters of our theoretical models i.e. ($\omega_0$, $\omega_1$) for both models A and B.
The 57 data points are composed of 31 points determined by the differential age method \cite{data3} and the remaining 26 points are determined by BAO measurements and other methods \cite{data2}. \\

 \textbf{Pantheon$^+$}:  we use the recent Pantheon Plus dataset  \cite{Pan}, which consists of 1701 light curves of 1550 distinct Supernova Type Ia (SNIa), in the redshift range $z \in [0.001, 2.26]$.  The chi-square function is written as
\begin{equation}
\chi_{SN}^2(\alpha,\beta)=\Delta\mu^T\mathcal{C}%
_{\text{Pan}^+}^{-1}\Delta\mu,
\end{equation}
where $\mathcal{C}_{\text{Pan}^+}$ is the covariance matrix and $\Delta\mu$ is a vector whose elements refer to the $i^{th}$ SNIa distance-modulus of the data set,  given by:
\begin{equation}
\Delta\mu^i=
\begin{cases}
\mu^i-\mu^i_{\text{ceph}}, &   i\in \text{ Cepheid host},\\
\mu^i-\mu_{\text{model}}(z_i\mid \alpha,\beta), &    \text{ Otherwise},
\end{cases}
\end{equation}
where $\mu^i_{\text{ceph}}$ is the distance modulus for the Cepheid host of the $i^{th}$ SNIa from SH0ES \cite{SH0ES} and  the theoretical distance modulus is defined as

\begin{equation}
\mu_{\text{model}}(z_i\mid \alpha,\beta)=5\log_{10}{\frac{d_L}{Mcp}}+25=m_b-M_B,
\end{equation}

where $m_b$ is the apparent magnitude, $M_B$ is the absolute magnitude, and $d_L$ is the luminosity distance, defined as
\begin{equation}
d_L=(1+z) c\int_{1}^{z}\frac{dz}{E(z)},
\end{equation}
with  $c$ is the speed of light, and $E(z)={H(z)}/{H_0}$ is the normalized Friedmann equation. \\

By combining the two datasets, the $\chi^2_{tot}$ is the sum of the chi-square of the $OHD$ data, $\chi^2_{OHD}$, and the one of the Pantheon$^+$ datasets, $\chi^2_{Pan+}$

\begin{equation}
\chi^2_{tot}=\chi^2_{OHD}+  \chi^2_{Pan+}.
\end{equation}

In order to determine which theoretical model provides a better description of the observational data, we use the Akaike's Information Criterion ($AIC$)  \cite{AIC}, which depends on the number of parameters, $N_p$\footnote{{$N_p$ = 3, 5 and 5 for $\Lambda$CDM, Model A and Model B, respectively.}}, given by

\begin{equation}
AIC = \chi^2_{min}+2\mathcal{N_{\textrm{p}}},
\end{equation}
we also define the $AIC$ corrected, $AIC_c$, which depends on the number of parameters, $N_p$ and the number of data points $N_{tot}$\footnote{$\mathcal{N_{\textrm{tot}}}=1758$ for OHD+Pantheon$^+$ datasets}, given by \cite{AIC1}

\begin{equation}
AIC_c = \chi^2_{min}+2\mathcal{N_{\textrm{p}}}+\frac{2\mathcal{N_{\textrm{p}}}(\mathcal{N_{\textrm{p}}}+1)}{\mathcal{N_{\textrm{tot}}}-\mathcal{N_{\textrm{p}}}-1},
\end{equation}

By definition, the model that has the lowest $AIC$ is considered to be a reference since it is most strongly supported by the observational data. In our analysis we take the standard model of cosmology, $\Lambda$CDM, as a reference model, where it is the most preferred model by several observation data. For this we calculate 
\begin{equation}
\Delta AIC =  AIC_{model}- AIC_{\Lambda CDM},
\end{equation}

in this case, a positive value of $\Delta AIC$ indicates that $\Lambda$CDM is the most preferred model, while a negative value of $\Delta AIC$  indicates that our model is preferred. Furthermore, if 0$\leqslant\Delta$AIC$<2$, this indicates that both models have approximately the same level of goodness of fit to the observational data and  if 4$\leqslant\Delta$AIC$<7$,  it suggests that the dataset slightly less supports the model, while for $\Delta $AIC$>10$, the dataset does not support the model.\\


\begin{table*}[!htp]
{\caption{The mean (best fit) values and the error at $1\protect\sigma$ (68\% C.L.) of the free parameters using OHD+ Pantheon+ dataset. Additionally, we show also $\chi^2_{tot}$, $\bigtriangleup AIC$ and  $\bigtriangleup BIC$.}\label{3}}
{
\centering
\begin{tabular}{c|c|c|c}
\hline
\hline
\multicolumn{1}{c|}{Data}    & \multicolumn{3}{c}{OHD+Pantheon$^+$}\\
\hline  
\multicolumn{1}{c|}{Model}  & \multicolumn{1}{c|}{ $\Lambda$CDM} & \multicolumn{1}{c|}{Model A}& \multicolumn{1}{c}{Model B} \\[0.1cm]
\hline
\multicolumn{1}{c|}{Parameters}    & \multicolumn{3}{c}{Mean $\pm 1\sigma$ (best-fit)}\\
\hline  
\hline 
$h$    & $0.71$ $\pm 0.0082$  &$ 0.7163\pm 0.0086$ ($ 0.7140$)   & $0.7154\pm 0.0086$  ($0.715$)
\\[0.1cm]
$\Omega_m$    & $ 0.282 \pm 0.01189$ &$ -$   &$-$-
\\[0.1cm]
$\lambda$    & $ -$& $-0.3339^{+0.0091}_{-0.013}$ ($-0.306$)   &$-0.3321^{+0.0088}_{-0.015}$ ($-0.316$)
\\[0.1cm]
$\omega_0$    & $ -$&$-1.192^{+0.071}_{-0.043}$ ($-1.107$)   &$-1.222^{+0.080}_{-0.053}$ ($-1.148$)
\\[0.1cm]
$\omega_1$    & $ -$&$-1.05^{+0.22}_{-0.33}$ ($-0.5034$)  &$-0.68^{+0.14}_{-0.26}$ ($-0.4501$)
\\[0.1cm]
$M_B$    & $-19.323\pm 0.0231$ &$-19.307\pm 0.024$ ($-19.312$)   &$-19.306\pm 0.025$ ($-19.306$)
\\[0.1cm]
\hline
\hline
$\chi^2_{tot}$   &$ 1564.111$&$1554.53$   &$1553.55$
\\[0.1cm]
$AIC$       &$1570.1$&$1564.6$ &$ 1563.6$
\\[0.1cm]
$\bigtriangleup AIC$       &$0$&$-5.5$ &$-6.5$
\\[0.1cm]
$BIC$      &$1586.52$&$ 1591.88$ &$ 1590.9$
\\[0.1cm]
$\bigtriangleup BIC$       &$0$&$ +5.36$ &$+4.38$
\\[0.1cm]
\hline       
\hline        
 
\end{tabular}
}
\end{table*}
\begin{figure}[h]
\centering
\centerline{\includegraphics[width=15cm,height=15cm]{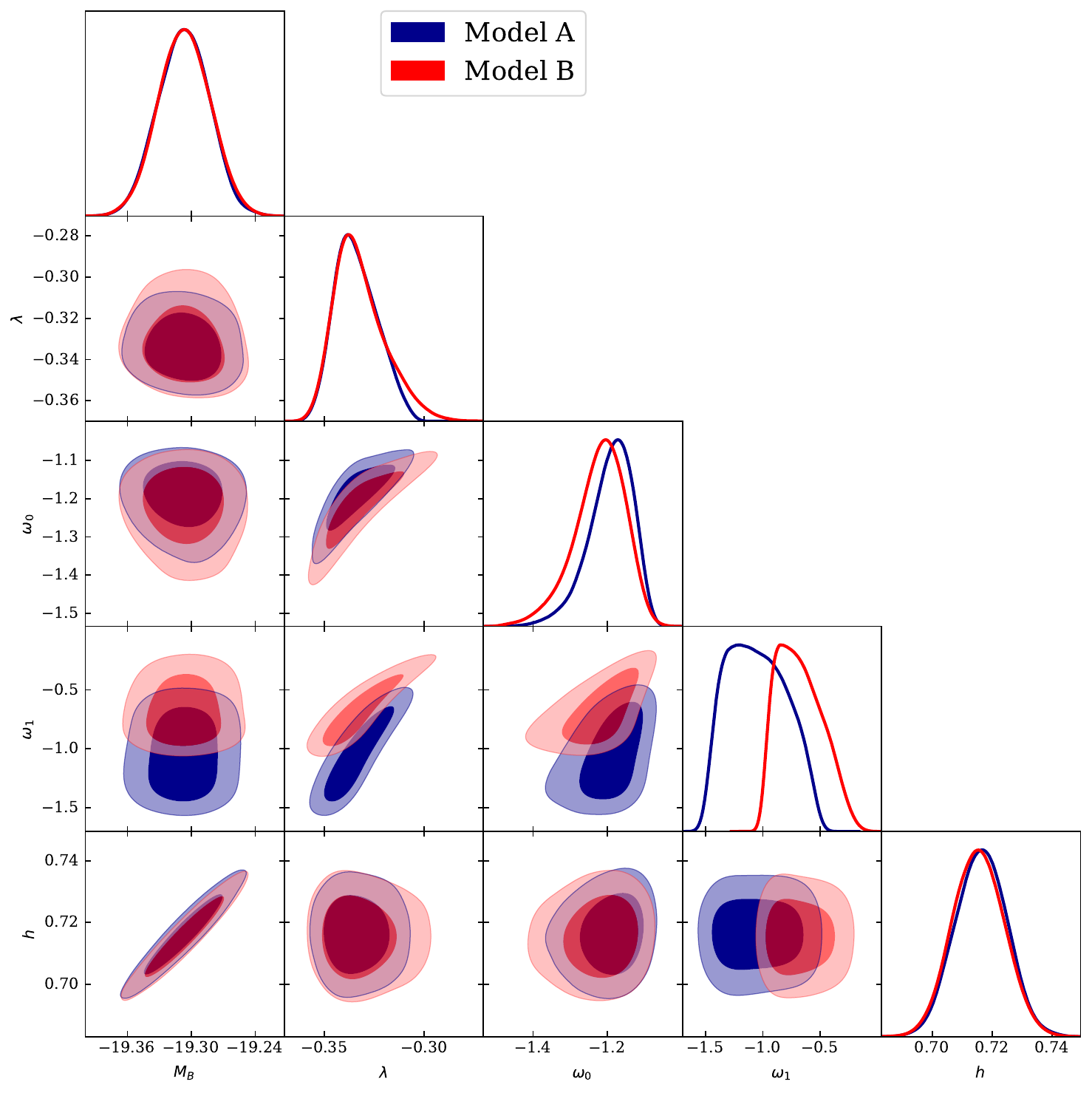} }
\caption{The 1D posterior distributions and  2D  confidence contours at $1\protect\sigma$ and $2%
\protect\sigma$ for A and B models using OHD+Pantheon$^+$ dataset.}
\label{f3}
\end{figure}

To complete our statistical study, we also use the selection method named the Bayesian Information Criterion (BIC) \cite{BIC}. To calculate the value of BIC for each model we use 
\begin{equation}
BIC = \chi^2_{min}+\mathcal{N_{\textrm{p}}}\ln{(\mathcal{N_{\textrm{tot}}})}.
\end{equation}
The difference $\Delta \textrm{BIC}$ with respect to $\Lambda$CDM, 
 is defined as
\begin{equation}
\Delta BIC =  BIC_{model}- BIC_{\Lambda CDM},
\end{equation}
as $\Delta AIC$, a positive value of $\Delta BIC$ indicates that our model is less preferred, while a negative value of $\Delta BIC$  indicates that our model is the more preferred by the observational data. In addition, 0$\leqslant\Delta$BIC$<2$ indicates that there is not enough evidence against the model with a large BIC value, 2$\leqslant\Delta$BIC$<6$ means that there is evidence and for 2$\leqslant\Delta$BIC$<10$ means that there is strong evidence against the model.
\\

\section{RESULTS AND DISCUSSIONS}\label{sec5}

In Table \ref{3}, we report the best fit and the mean$\pm 1\sigma$ of the cosmological parameters \{$h$, $\Omega_m$, $M_B$\} for the $\Lambda$CDM model and \{$h$, $\lambda$, $\omega_0$, $\omega_1$, $M_B$\} for models A and B. Using the combination of OHD and Pantheon$^+$ datasets where $H_0=100$ $h$ kms$^{-1}$ Mpc$^{-1}$,  and $M_B$  the absolute magnitude and $\lambda$, $\omega_0$ and $\omega_1$ are the parameters that characterize the  models A and B. Model selection criteria $\Delta$AIC and $\Delta$BIC are shown in the same Table. In Fig. \ref{f3}, we show  1D and 2D posterior distributions at 1$\sigma$ and 2$\sigma$  for the cosmological parameters using Getdist \cite{GET}.\\

By constraining the $\Lambda$CDM, A and B models with OHD+Pantheon$^+$ dataset, we get $h=0.71\pm 0.0082$ and $M_B=-19.323\pm 0.0231$, for $\Lambda$CDM. A small increase for the current Hubble parameter is observed for the model A and the model B. In this context, we obtain $h=0.7163\pm 0.0086$ ($h=0.7154\pm 0.0086$) for the model A (model B).  We find also a small increase for $M_B$, i.e. $M_B=-19.307\pm 0.024$ ($M_B=-19.306\pm 0.025$) for the model A (model B) (see also Fig. \ref{f2} and Fig. \ref{f1}). In the case of CPL  parameterization i.e. model A, the mean value of the EoS parameter are $\omega_{0}=-1.192^{+0.071}_{-0.043}$ and $\omega_{1}=-1.05^{+0.22}_{-0.33}$. While, in the logarithmic parametrization i.e. model B, we obtain $\omega_{0}=-1.222^{+0.080}_{-0.053}$ and $\omega_{1}=-0.68^{+0.14}_{-0.26}$. We also observe that $\lambda$ is positively correlated with $\omega_0$ and $\omega_{1}$ for both models (see Fig. \ref{f3}). As the EoS parameters for both models are almost equal, $\lambda$ has also almost the same values for both of them. The best fit of the EoS parameter, take the values $\omega_{0} =-1.107\pm 0.0619$ and $\omega_{1}=-0.5034 \pm 0.2531$ which is close to $\Lambda$CDM with $\omega_{0} =-1.01$  \cite{Planck} at $z=0$ but mimic the phantom behaviour.\\

On the other hand, we note that models A and B have a minimum value of $\chi^2$ compared to the $\Lambda$CDM model, with a difference of $\sim 10$. This indicates that our models demonstrate a strong fit to OHD+Pantheon$^+$ datasets.  This result may be due to the additional parameters. However, $\chi^2$ cannot be considered as an optimal criterion to select the best models, due to different number of cosmological parameters.  Therefore, we have calculated  AIC and BIC.  According to Table \ref{3}, we see that the small values of AIC are obtained for models A and B,  and that  4$\leqslant\lvert\Delta$AIC$\lvert<7$ this indicates that both models are well supported by the observational data than the $\Lambda$CDM model. On the other hand, for the BIC criterion, we obtained a positive value for $\Delta$BIC with 2$\leqslant\lvert\Delta$BIC$\lvert<10$, which means that our models are less supported by the observational data. This result is expected as the BIC criterion strongly penalizes models with additional parameters. \\
In Fig. \ref{fmu}, we present the evolution of $H(z)$ (left panel) and of $\mu(z)$ (right panel) for  each models, using the best-fit values of the parameters obtained in Table \ref{3}. We see that the evolution of H(z) and $\mu(z)$ for the two models A and B behaves similarly to the $\Lambda$CDM model. \\
From Fig. \ref{decelerationq}, it is evident that the deceleration parameter accounts for two
phases of the Universe, i.e. the phases where the Universe experiences a deceleration expansion and the ones where it accelerates.
Such behavior is commonly observed in several works. The transition between these two phases
occurred at  the transition redshift values $z_{i}=0.605$, 
and $z_{i}=0.605$ for the model $A$, and $B$, respectively. Furthermore, the current values of the deceleration parameter are $q_{0}=-0.5$, and $q_{0}=-0.52$ for the model $A$, and $B$, respectively. These negative values agree with the observed acceleration phase of the Universe.

\begin{figure}[h]
\centering
\centerline{\includegraphics[width=12cm,height=8cm]{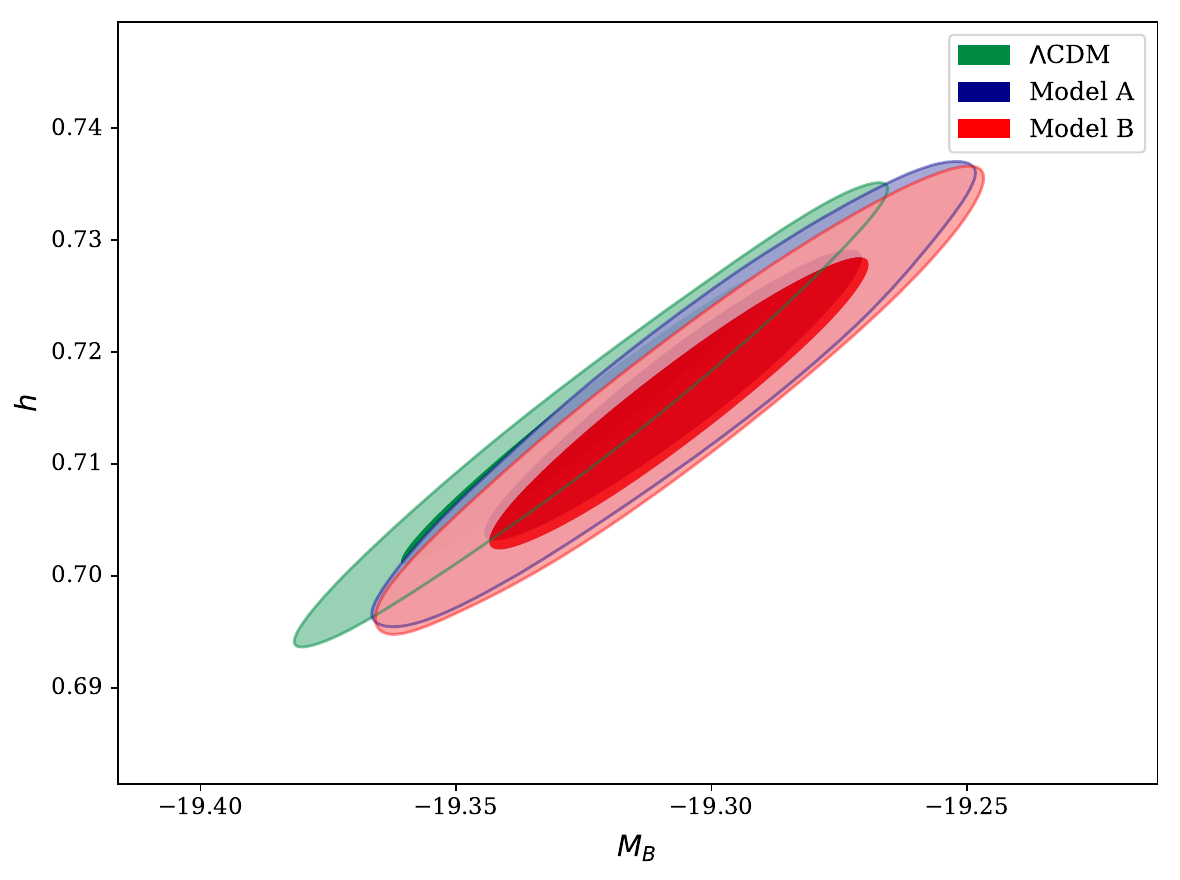} }
\caption{The \{h,$M_B$\} contour at $1\protect\sigma$ and $2 \protect\sigma$ for $\Lambda$CDM (in green), model A (in blue) and model B (in red), using OHD+Pantheon$^+$ dataset.}
\label{f2}
\end{figure}
\begin{figure}[h]
\centering
\centerline{\includegraphics[width=15cm,height=7cm]{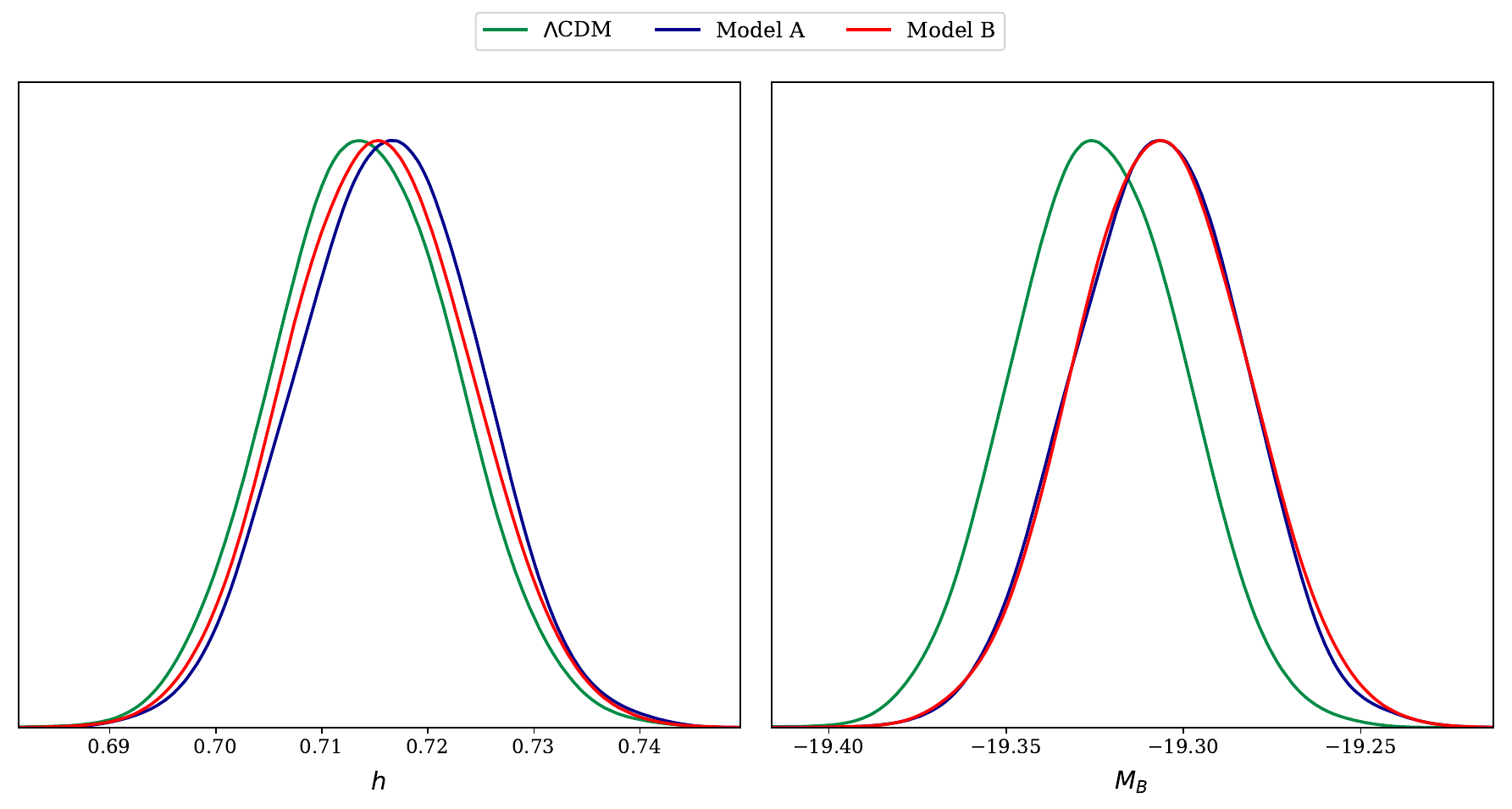} }
\caption{The 1D posterior distributions for $h$ (left panel) and $M_B$ (right panel) for $\Lambda$CDM (in green), model A (in blue) and model B (in red).}
\label{f1}
\end{figure}

\begin{figure*}
\centering
\includegraphics[width=8.7cm,height=7cm]{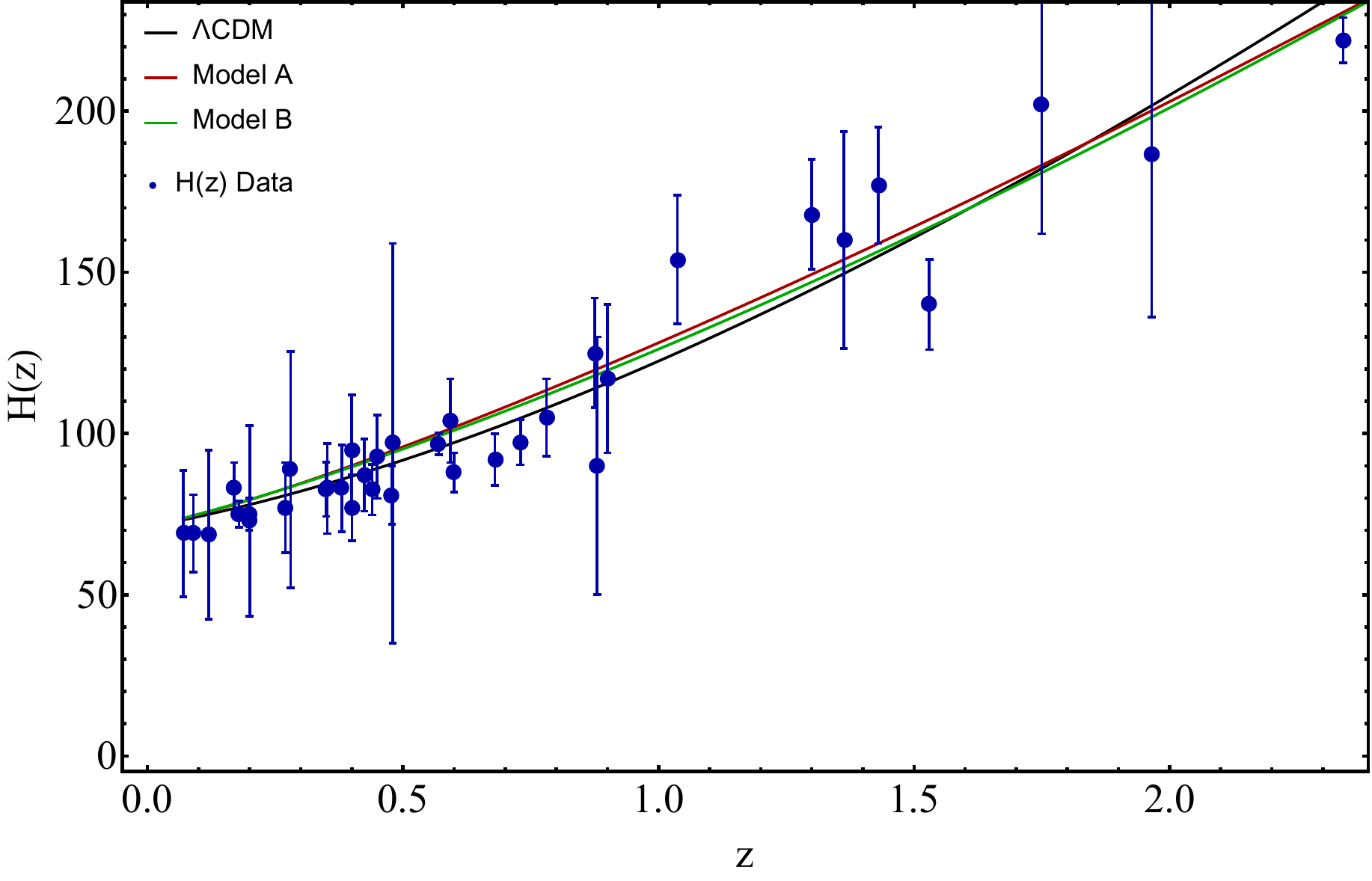}\hfill
\includegraphics[width=8.7cm,height=7cm]{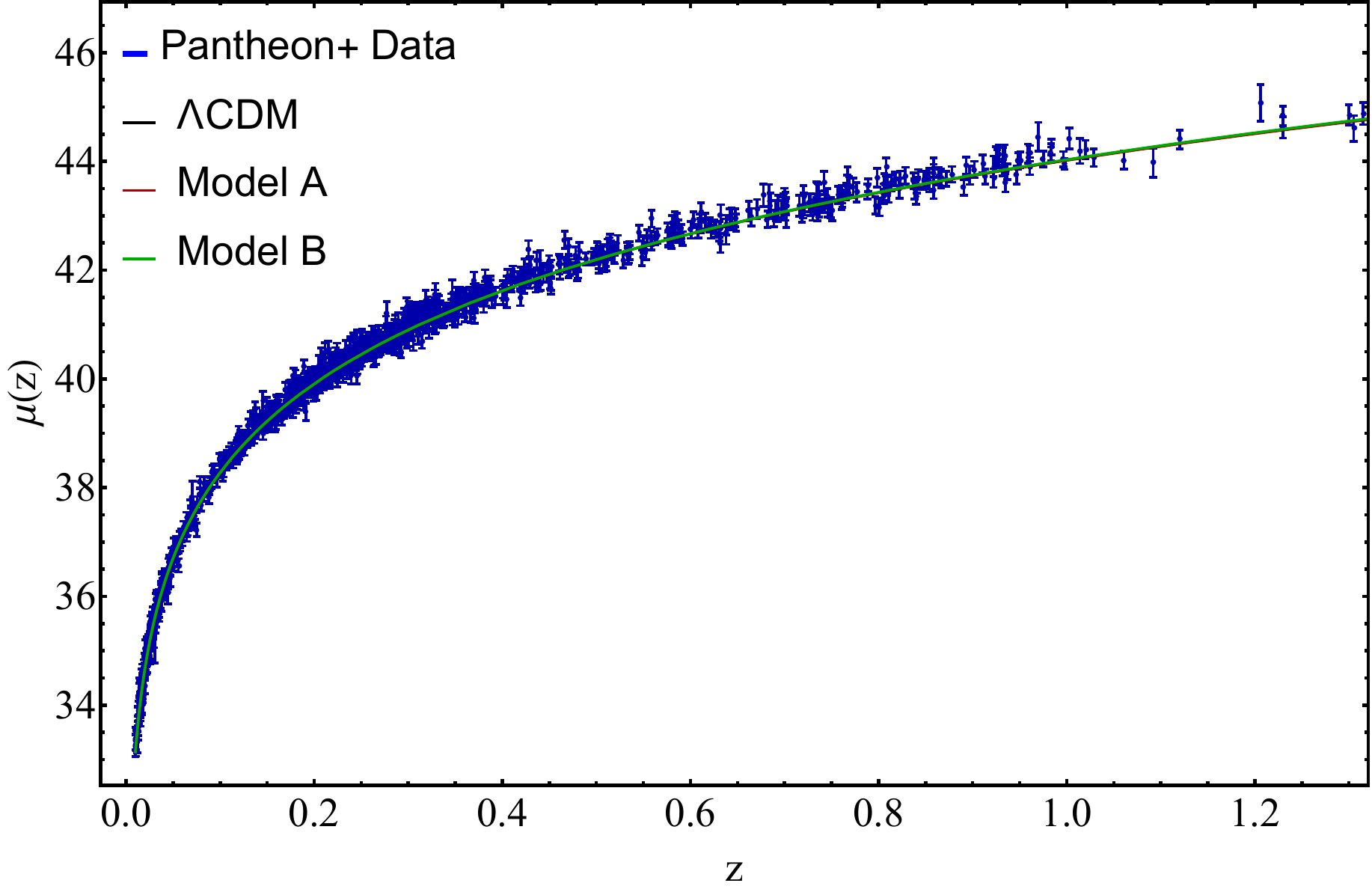}
\caption{The figure shows the evolution of Hubble parameter $H(z)$ (left panel) and the evolution of $\mu(z)$ (right panel), for $\Lambda$CDM (black color), model A (magenta color) and model B (green color) using the best-fit values from Tab \ref{3}}.
\label{fmu}
\end{figure*}

\begin{figure}[h]
\centerline{\includegraphics[scale=0.3]{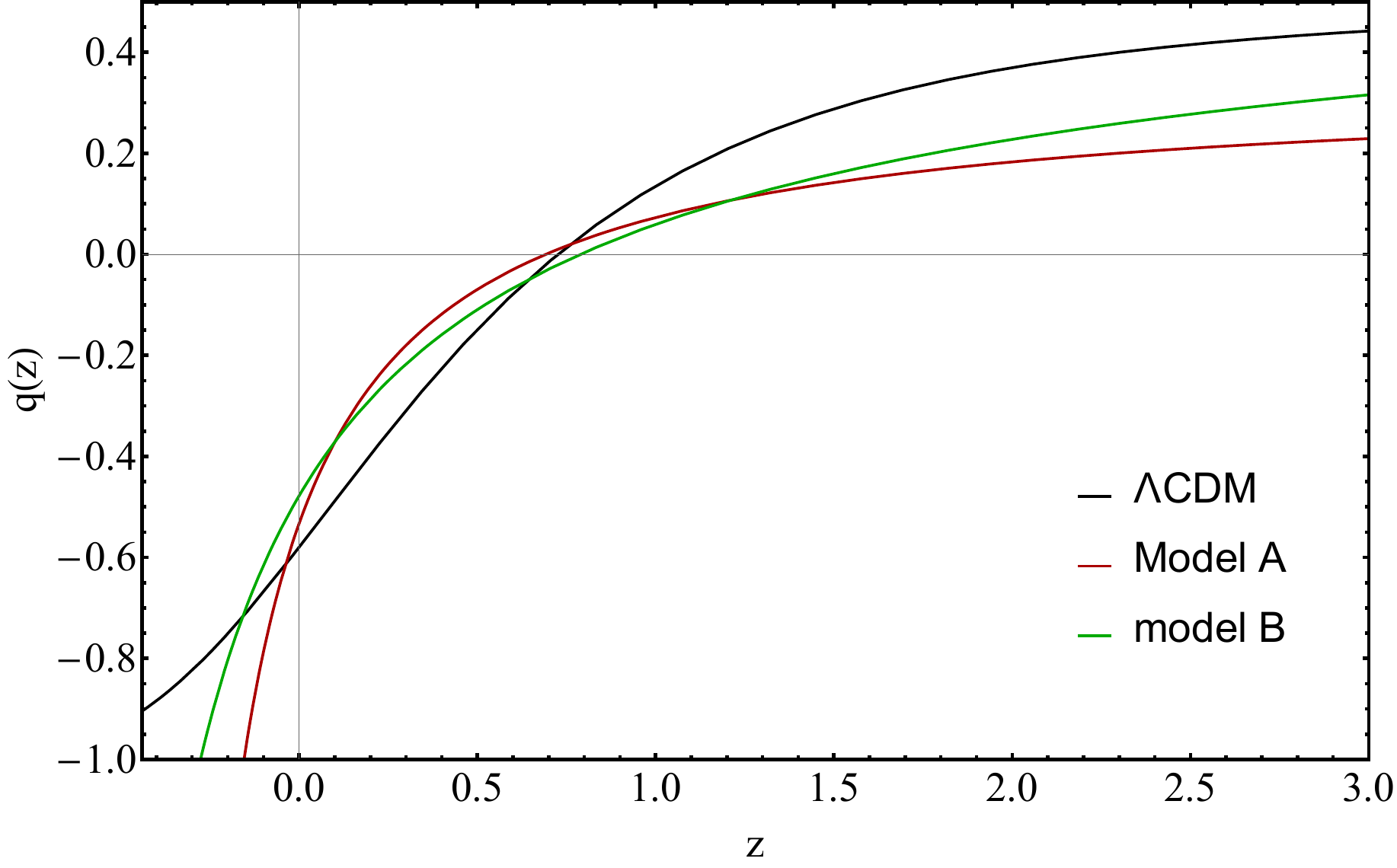}}
\caption{The behavior of the deceleration parameter in terms of redshift $z$ using  the best-fitting value indicated in Tab \ref{3}}.
\label{decelerationq}
\end{figure}


 
 \section{Conclusions}\label{sec6}
\label{sec6}

 In this study, we have discussed the late-time acceleration of the Universe within the framework of $F(R,T)$ gravity, where $R$ denotes the Ricci scalar and $T$ represents the trace of the energy-momentum tensor. Our analysis is focused on the specific case where the gravitational action is given by $f(R,T)=R+2\kappa^{2}\lambda T$, with $\lambda$ being a constant parameter. We have considered specifics dark energy model sourced from a geometric concept of F(R,T) gravity. 
To parametrize dark energy, we considered two well known equations of state: Model $A$, represented by the Chevallier-Polarski-Linder  parametrization, and Model $B$, characterized by a logarithmic form of the equation of state parameter.
Utilizing observational data including 57 data points of $H(z)$ measurements (OHD) and 1701 data points of the Pantheon+ dataset, we constrained cosmological parameters, particularly the Hubble parameter h and the Supernova absolute magnitude MB. We have obtained the mean and the best fit values of each model by performing a Markov Chain Monte Carlo analysis. 
Firstly, we observed a slight increase in the Hubble parameter and the SN absolute magnitude for both models compared to $\Lambda$CDM. Additionally, the minimum value of the $\chi^2$ function favored both models $A$ and $B$ over the $\Lambda$CDM model. Moreover, in order to classify these two models, we considered the AIC and the BIC information criterion by means of the degrees of freedom of the two models and the total number of the data points used by both combinations. Although the AIC criterion supports models A and B against the $\Lambda$CDM,  these models are less supported by the BIC , as this criterion is known to penalize models with additional parameters. 
Lastly, the evolution of the Hubble parameter $H(z)$ and the distance modulus $\mu(z)$ with redshift exhibite a similar behavior for the two models and $\Lambda$CDM.
These findings collectively enhance our understanding of the behavior of dark energy within the context of $F(R,T)$ gravity and its implications for cosmological dynamics.
Furthermore, our analysis revealed that the Universe recently underwent a transition from deceleration to acceleration for both models $A$ and $B$. Additionally, the equation of state parameter exhibite a phantom nature of dark energy.
While our study provides valuable insights into the behavior of dark energy within the context of $F(R,T)$ gravity, it is essential to acknowledge some limitations inherent in this framework, including theoretical complexities and challenges associated with specific forms of $F(R,T)$ gravity.
Looking ahead, future research directions may involve exploring alternative forms of $F(R,T)$ gravity and investigating the compatibility of our results with other dark energy models, such as phantom dark energy models. By further refining our understanding of the interplay between modified gravity theories and dark energy, we can continue to deepen our comprehension of fundamental properties driving the dynamics of the Universe.
In conclusion, our study contributes to advancing the understanding of cosmological evolution within the framework of $F(R,T)$ gravity and provides a basis for further exploration of alternative gravity theories  and dark energy models. Furthermore, our results have shown that the geometric part of $F(R,T)$ gravity may be considered as an alternatives to dark energy.


\end{document}